\documentclass[conference]{IEEEtran}
\usepackage{paper}
\IEEEoverridecommandlockouts

\definecolor{commentcolor}{RGB}{110,154,155}   
\def\BibTeX{{\rm B\kern-.05em{\sc i\kern-.025em b}\kern-.08em
    T\kern-.1667em\lower.7ex\hbox{E}\kern-.125emX}}

\begin{document}
\title{RSSI-Assisted CSI-Based Passenger Counting with Multiple Wi-Fi Receivers}

\author[]{Jingtao Guo, Wenhao Zhuang, Yuyi Mao, and Ivan Wang-Hei Ho \\ Dept. of EEE, The Hong Kong Polytechnic University, Hong Kong \\ jingtao2023.guo@connect.polyu.hk, \{wzhuan, yuyi-eie.mao, ivanwh.ho\}@polyu.edu.hk}
\maketitle              
\begin{abstract}
Passenger counting is crucial for public transport vehicle scheduling and traffic capacity evaluation. However, most existing methods are either costly or with low counting accuracy, leading to the recent use of Wi-Fi signals for this purpose. In this paper, we develop an efficient edge computing-based passenger counting system consists of multiple Wi-Fi receivers and an edge server. It leverages channel state information (CSI) and received signal strength indicator (RSSI) to facilitate the collaboration among multiple receivers. Specifically, we design a novel CSI feature fusion module called Adaptive RSSI-weighted CSI Feature Concatenation, which integrates locally extracted CSI and RSSI features from multiple receivers for information fusion at the edge server. Performance of our proposed system is evaluated using a real-world dataset collected from a double-decker bus in Hong Kong, with up to 20 passengers. The experimental results reveal that our system achieves an average accuracy and F1-score of over 94\%, surpassing other cooperative sensing baselines by at least 2.27\% in accuracy and 2.34\% in F1-score.
\end{abstract}

\section{Introduction}
Urban commuting has surged recently across various transport modes, including private cars, taxis, buses, and subways. Real-time passenger data is vital for stakeholders like automobile manufacturers, transit operators, and urban planners. For private vehicles, it informs product optimization and improves dispatch efficiency, reducing wait times. In public transit, passenger occupancy data supports real-time scheduling and resource allocation, enhancing service quality and aiding sustainable urban development~\cite{kumar2018fast}.

Crowd-counting methods can be broadly categorized into device-based and device-free approaches~\cite{pcsurvey}. Wi-Fi-based techniques are notable for their ability to perform contactless, non-line-of-sight (NLOS) counting using existing Wi-Fi infrastructure, making them cost-effective and practical for indoor environments without the need for additional sensors or cameras. As Wi-Fi signals scatter and reflect based on the number of people present, they serve as a reliable tool for crowd counting and occupancy detection~\cite{pcsurvey}. Key data derived from Wi-Fi signals, such as received signal strength indicator (RSSI) and channel state information (CSI) are essential in these tasks. While RSSI-based methods are effective in certain scenarios~\cite{depatla2015occupancy}, they lack robustness and scalability due to multipath fading effects~\cite{yang2013rssi}. In contrast, CSI provides more detailed data, including amplitude and phase information across multiple sub-carriers, making it more suitable for accurate crowd counting~\cite{wifivision, guo2023fedpos}.

Several studies have utilized CSI-based occupancy measurement systems with a single pair of transceivers for crowd sensing across various scenarios~\cite{guo2022csi},\cite{jiang2023pa}. However, indoor environments with multiple reflectors can cause deep fading that distorts CSI measurements\cite{bahadori2022rewis}, reducing accuracy and coverage in complex settings. Modern Wi-Fi setups use multiple access points (APs) to extend range, improve reliability, and support numerous devices, emphasizing the need for better collaboration among Wi-Fi sensors to harness signal diversity from different locations. For instance, \cite{yin2022fewsense} averaged probability distributions from multiple receivers to generate the final prediction; however, it did not fully utilize the CSI data collected from different receivers. This paper addresses this gap by proposing a new approach that improves cooperative sensing through integrating CSI and RSSI data from multiple positions. Specifically, it introduces a CSI feature fusion module inspired by the relationship between RSSI and CSI\cite{yang2013rssi}, enhancing sensing accuracy and F1-score compared to single or multiple receiver cases. The main contributions of this paper are three-fold:
\begin{itemize}
    \item We introduce a system that uses multiple WiFi receivers for cooperative sensing to enhance crowd counting capabilities.
    \item We propose an RSSI-Aided CSI feature fusion module that integrates CSI features from multiple receivers and utilizes RSSI to enhance the fused CSI feature. This approach also demonstrates higher computational efficiency compared to methods that rely on probability ensembles.
    \item We evaluate our system in a real-world scenario on a double-decker bus with up to 20 passengers. Results demonstrate that our system achieves an average accuracy and F1-score above 94\%, with improvements of at least 2.27\% in accuracy and 2.34\% in F1-score compared to other cooperative sensing baselines.
\end{itemize}

This paper is organized as follows. Section \uppercase\expandafter{\romannumeral2} introduces the preliminaries, including CSI background and data preprocessing. Section \uppercase\expandafter{\romannumeral3} presents the proposed system, focusing on the system overview and the RSSI-assisted CSI feature fusion approach. Section \uppercase\expandafter{\romannumeral4} discusses the experimental setup and results, and Section \uppercase\expandafter{\romannumeral5} concludes with a summary of key findings and contributions.

\section{Preliminaries}
In this section, we present the preliminaries of Wi-Fi CSI, along with the data preprocessing steps for Wi-Fi CSI data.
\subsection{Wi-Fi CSI}
Wi-Fi technologies utilize orthogonal frequency division multiplexing (OFDM), where data and reference symbols are transmitted via orthogonal subcarriers. Wi-Fi CSI represents the estimated channel response between a Wi-Fi transmitter (Tx) and a receiver (Rx) for each subcarrier. This information is readily available at the receiver and can be extracted using CSI extraction middlewares such as Nexmon CSI extractor~\cite{gringoli2019free}.

Assume that $K$ data packets, each with $S$ subcarriers, are transmitted between $N$ single-antenna Wi-Fi Tx-Rx pairs, we denote the Wi-Fi CSI data collected via the $n$-th Wi-Fi Tx-Rx pair as $H_n \in \mathbb{C}^{K \times S}$, where the $(k, s)$-th entry of $H_n$ represents the channel response between the $n$-th antenna pair for the $k$-th packet at the $s$-th subcarrier as follows:
\begin{equation}
\label{eqn_1}
{H}_{n}[k, s] = {A}_{n}[k, s]e^{j \cdot \theta_{n}[k, s]},
\end{equation}
where ${A}_{n}[k, s]$ and $\theta_{n}[k, s]$ denote the amplitude and phase responses, respectively. 
Since the phase responses may suffer from severe distortions due to carrier frequency offset and sampling timing offset~\cite{tewes2021ws}, the amplitude responses are more commonly used for deep learning-based Wi-Fi sensing applications. 
Therefore, we utilize amplitude responses in this system to streamline data preprocessing. Besides, we denote $y$ as the label of each Wi-Fi CSI sample, indicating the number of people in the environment of interest.

\subsection{Wi-Fi CSI Pre-processing}
\label{sec:datapreprocessing}
Wi-Fi CSI data pre-processing is required to enhance the signal quality for improving the performance of various sensing applications. It typically includes three steps, namely outlier removal and noise reduction, CSI rescaling, and data segmentation. 

\begin{enumerate}
    \item \textbf{Outlier Removal and Noise Reduction:}
    Null subcarriers, i.e., subcarriers that carry no data symbols, can introduce outliers and degrade learning performance~\cite{gast2013802}.
    Then, a Savitzky-Golay filter (SG filter), i.e., a least-square smoothing filter that fits successive data point subsets with a low-degree polynomial using linear least squares, is employed to further mitigate small random fluctuations while preserving significant activity changes.
    \item \textbf{CSI Rescaling:} Since the automatic gain control (AGC) module in a Wi-Fi receiver scales received signals into relative values, the extracted CSI data does not accurately reflect the true channel responses~\cite{wei2022rssi}. Therefore, rescaling the Wi-Fi CSI data using the AGC factor is necessary. However, as the AGC factor may not always be accurately available, we apply the RSSI-assisted CSI rescaling method proposed in~\cite{gao2020crisloc}, as described below:
    \begin{equation}
        \label{eqn_3}
        \tilde{A}_{n}[k, s] = {A}_{n}[k, s] \cdot \sqrt{\frac{10^{\text{RSSI}_{n}[k]}/10}{\sum_{i=1}^{S} ( {A}_{n}[k, s])^2}},
    \end{equation}
    where $\text{RSSI}_{n}[k]$ (dBm) denotes the received signal power of the $k$-th packet for the $n$-th Wi-Fi Tx-Rx pair. 

    \item \textbf{Data Segmentation:} Window cropping, a basic data augmentation technique for time series data~\cite{wen2020time}, can be applied to Wi-Fi CSI data. Amplitude responses are extracted from the rescaled CSI data and a sliding window $\mathbf{W}\in \mathbb{R}^{T_{w} \times S}$, where all elements are set to $1$, is applied to slide along the first dimension with stride $T_s$ and produce CSI data segments as ${\mathbf{D}}_{n, p} \in \mathbb{R}^{T_{w} \times S}, n=1, \cdots, N, p=1, \cdots, \lceil \frac{K-T_w}{T_s} \rceil$, where $\lceil \cdot \rceil$ denotes the ceiling function, and $N$ is the total number of Wi-Fi Tx-Rx pairs. To increase the amount of training samples and ensures comprehensive coverage of CSI variations due to human fidgeting over different time periods, an overlapping segmentation approach is adopted by configuring $T_s < T_w$. 
\end{enumerate}

\begin{figure}[t]
    \centering
    \includegraphics[width=1.0\linewidth]{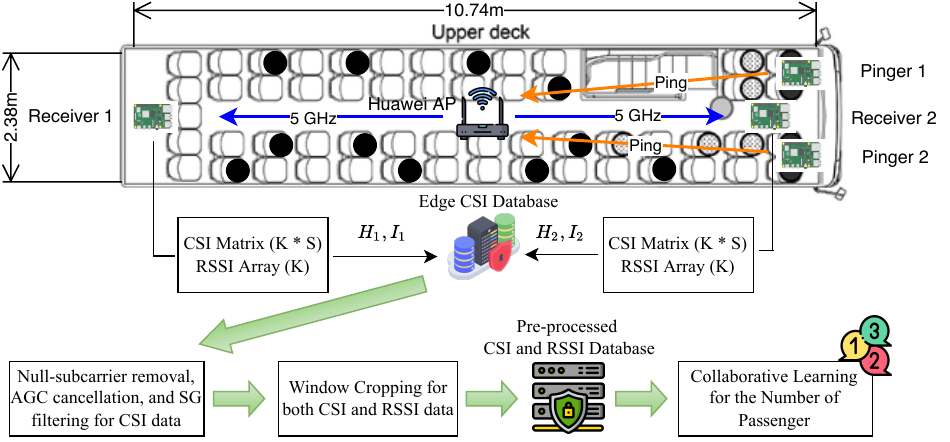}
    \caption{Overview of our proposed RSSI-assisted CSI-based cooperative sensing system with two Wi-Fi receivers.}
    \label{fig:pic1}
\end{figure}
\section{Proposed Framework}
In this section, we begin by introducing the proposed CSI-based passenger counting system utilizing multiple Wi-Fi receivers. Following that, we propose a RSSI-assisted CSI feature fusion method to enable cooperative sensing across multiple receivers, improving counting accuracy.

\subsection{System Overview}
Fig.~\ref{fig:pic1} illustrates the proposed system, where two single-antenna Wi-Fi receivers (Pinger 1, Pinger 2) send control messages to the Wi-Fi AP to generate data traffic, while two additional single-antenna Raspberry Pis (Receiver 1, Receiver 2) function as Wi-Fi receivers to collect CSI data. The collected data undergo the pre-processing steps outlined in Sec. \ref{sec:datapreprocessing}. The pre-processed CSI data from different receivers are then used for cooperative passenger counting.

\subsection{RSSI-Assisted cooperative Passenger Counting}
\begin{figure*}[t]
    \centering
    \includegraphics[width=0.9\linewidth]{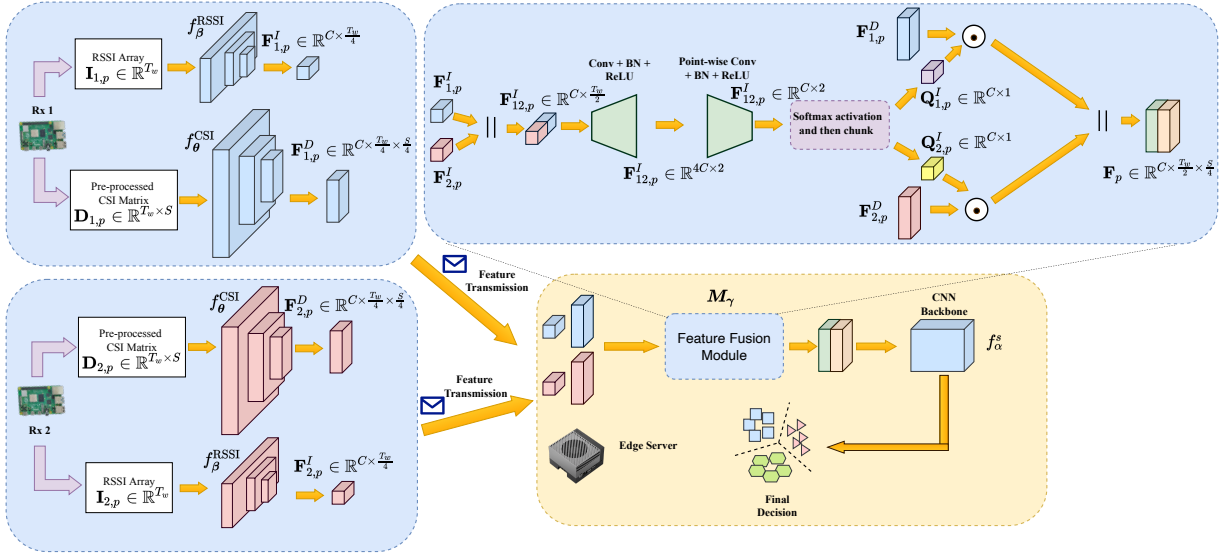}
    \caption{A two-receivers example workflow of our proposed cooperative sensing approach.}
    \label{fig:pic6}
\end{figure*}

\begin{table}[htbp]
\centering
\caption{Model architecture of the feature extractors.}
\label{tab:tab1}
\resizebox{1.0\linewidth}{!}{
\begin{tabular}{cc|cc}
\toprule 
\multicolumn{2}{c|}{\textbf{CSI Feature Extractor}} & \multicolumn{2}{c}{\textbf{RSSI Feature Extractor}} \\ \midrule
\textbf{Layers} & \textbf{Details} & \textbf{Layers} & \textbf{Details} \\ 
\multirow{2}{*}{1} & Conv2D(1, 64, 3, 2, 1) & \multirow{2}{*}{1} & Conv1D(1, 64, 3, 2, 1) \\
{} & BN2D(64), ReLU & {} & BN1D(64), ReLU \\
\multirow{2}{*}{2} & Conv2D(64, 128, 1, 1, 0) & \multirow{2}{*}{2} & Conv1D(64, 128, 1, 1, 0) \\
{} & BN2D(128), ReLU & {} & BN1D(128), ReLU \\
\multirow{2}{*}{3} & Conv2D(128, 64, 3, 2, 1) & \multirow{2}{*}{3} & Conv1D(128, 64, 3, 2, 1) \\
{} & BN2D(64), ReLU & {} & BN1D(64), ReLU \\
\bottomrule
\end{tabular}
}
\end{table}
Given the pre-processed CSI data from different receivers, it is crucial to develop an effective method to utilize the vast information among the CSI data. The proposed RSSI-assisted, CSI-based passenger counting system with multiple Wi-Fi receivers is depicted in Fig.~\ref{fig:pic6}. In this system, a cooperative sensing method is developed to incorporate additional sensing information from RSSI into the CSI data. Specifically, a RSSI-based feature fusion module is introduced with the details of each part provided below.

\subsubsection{Feature Extractor}
CSI data and RSSI data collected via each Wi-Fi receiver are utilized for counting. Since these raw data are rich in passenger fidgeting and environmental information but redundant and unstructured, the receivers exploit feature extractors to transform the raw data into compact and informative features. Denote $\mathbf{I}_{n, p} \in \mathbb{R}^{T_w}$ as the RSSI data collected via the $n$-th Wi-Fi Tx-Rx pair for the $p$-th CSI segment. Besides, we use $f^{{\text{CSI}}}_{\boldsymbol{\theta}}(\cdot)$ and $f^{{\text{RSSI}}}_{\boldsymbol{\beta}}(\cdot)$ to denote the 2D and 1D three-layer convolutional neural network (CNN) with trainable parameters $\boldsymbol{\theta}$ and $\boldsymbol{\beta}$, and the network architectures are given in Table~\ref{tab:tab1}, where we list parameters with sequence of input and output dimension, kernel size, stride and padding for convolutional layer (Conv2D and Conv1D), and the channel dimension for batch normalization layer (BN2D and BN1D). The feature extraction process can be expressed as
\begin{align}
    \mathbf{F}^D_{n, p} &\triangleq f^{{\text{CSI}}}_{\boldsymbol{\theta}}(\mathbf{D}_{n, p}), \\
    \mathbf{F}^I_{n, p} &\triangleq f^{{\text{RSSI}}}_{\boldsymbol{\beta}}(\mathbf{I}_{n, p}),
\end{align}
where $\mathbf{F}^D_{n, p}  \in \mathbb{R}^{C \times \frac{T_w}{4} \times \frac{S}{4}}$ and $\mathbf{F}^I_{n, p}  \in \mathbb{R}^{C \times \frac{T_w}{4}}$ are the extracted features from the raw CSI and RSSI data respectively, and $C$ denotes as the number of feature maps. \begin{table}[htbp]
\centering
\caption{Model architecture of the feature fusion module.}
\begin{tabular}{cc}
\toprule 
\textbf{Layers} & \textbf{Details}\\ \midrule
\multirow{2}{*}{1} & Conv1D(64, 256, $\frac{T_w}{4}$, $\frac{T_w}{4}$, 0) \\
& BN1D(256), ReLU \\
\multirow{2}{*}{2} & Conv1D(256, 64, 1, 1, 0) \\
& BN1D(64), Softmax, Chunk \\
\bottomrule
\end{tabular}
\label{tab:tab2}
\end{table}Then, the informative CSI and RSSI features are transmitted to the edge server for further processing, which can also reduce the communication cost compared with raw data transmission.

\subsubsection{RSSI-Assisted Feature Fusion}
We assume that the RSSI can effectively represents signal quality under identical receiver configurations. Although other metrics, such as the signal-to-noise ratio (SNR), might offer a more accurate assessment, these alternatives will be considered in future work. High RSSI readings from a receiver usually suggest a strong signal, which can therefore lend greater importance to the CSI measurements gathered under these conditions. Accordingly, we propose a novel RSSI-based feature fusion module to enhance the fused CSI features by leveraging the additional sensing information provided by RSSI data. Specifically, the RSSI weight is first generated from $\mathbf{F}^I_{n, p}$ and then multiplied to $\mathbf{F}^D_{n, p}$ for feature fusion. Denote $\boldsymbol{M}_{\boldsymbol{\gamma}}(\cdot)$ as the proposed RSSI weight generation module with trainable parameters $\boldsymbol{\gamma}$ and the module architecture is given in Table~\ref{tab:tab2}. Consider a two-receiver scenario, we can obtain $\mathbf{Q}^I_{1, p} \in \mathbb{R}^{C \times 1}$ and $\mathbf{Q}^I_{2, p} \in \mathbb{R}^{C \times 1}$, which are the RSSI weights for $\mathbf{F}^D_{1, p}$ and $\mathbf{F}^D_{2, p}$, respectively, as follows:  
\begin{align}
    \left[ \mathbf{Q}^I_{1, p}, \mathbf{Q}^I_{2, p} \right] = \boldsymbol{M}_{\boldsymbol{\gamma}}(\mathbf{F}^I_{1, p}, \mathbf{F}^I_{2, p}).
\end{align}
\begin{figure}[htbp]
    \centering
    \begin{subfigure}[htbp]{0.49\linewidth}
        \includegraphics[width=\linewidth]{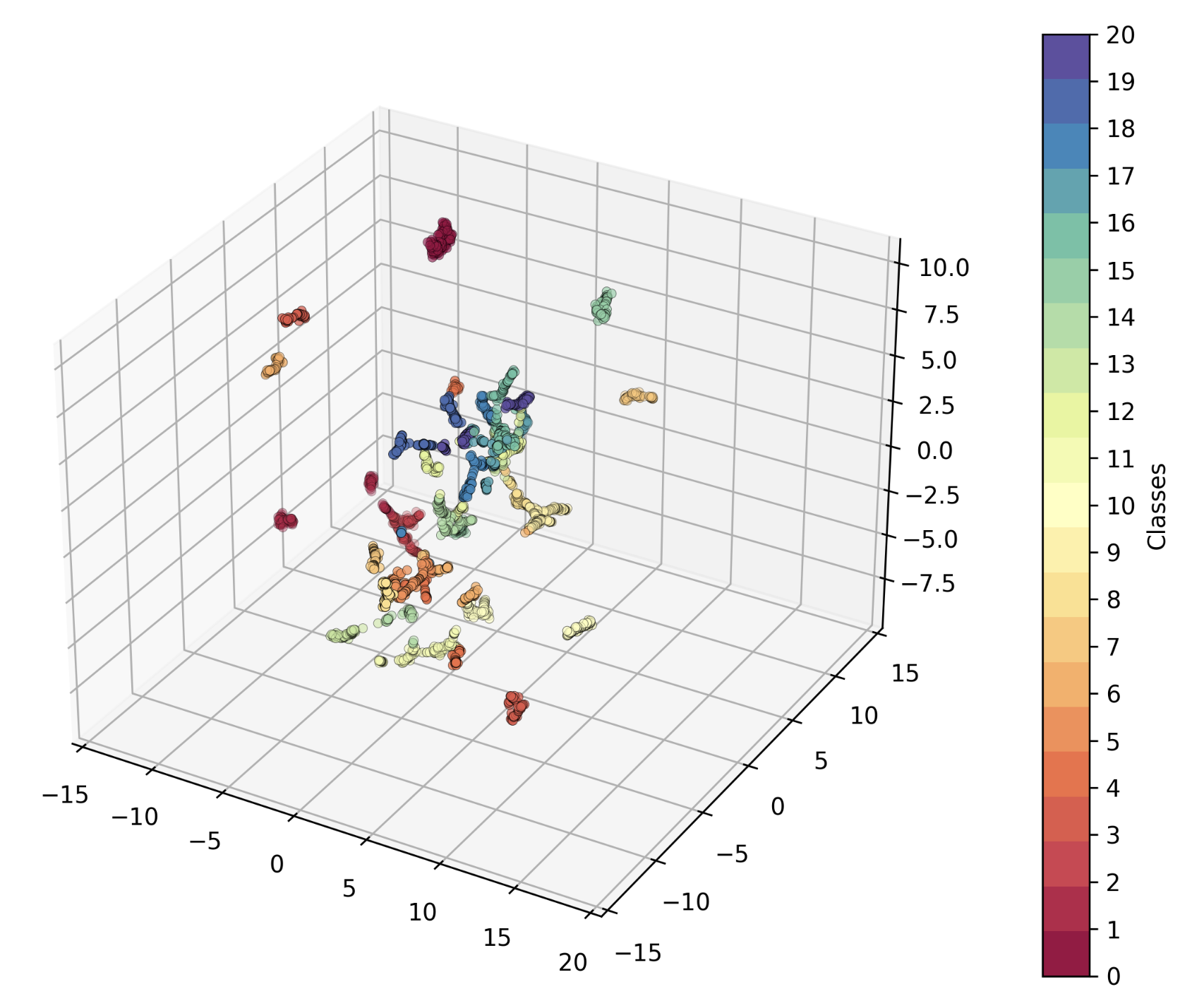}
        \caption{RSSI-weighted concatenated CSI features}
        \label{fig:img5}
    \end{subfigure}
    \begin{subfigure}[htbp]{0.49\linewidth}
        \includegraphics[width=\linewidth]{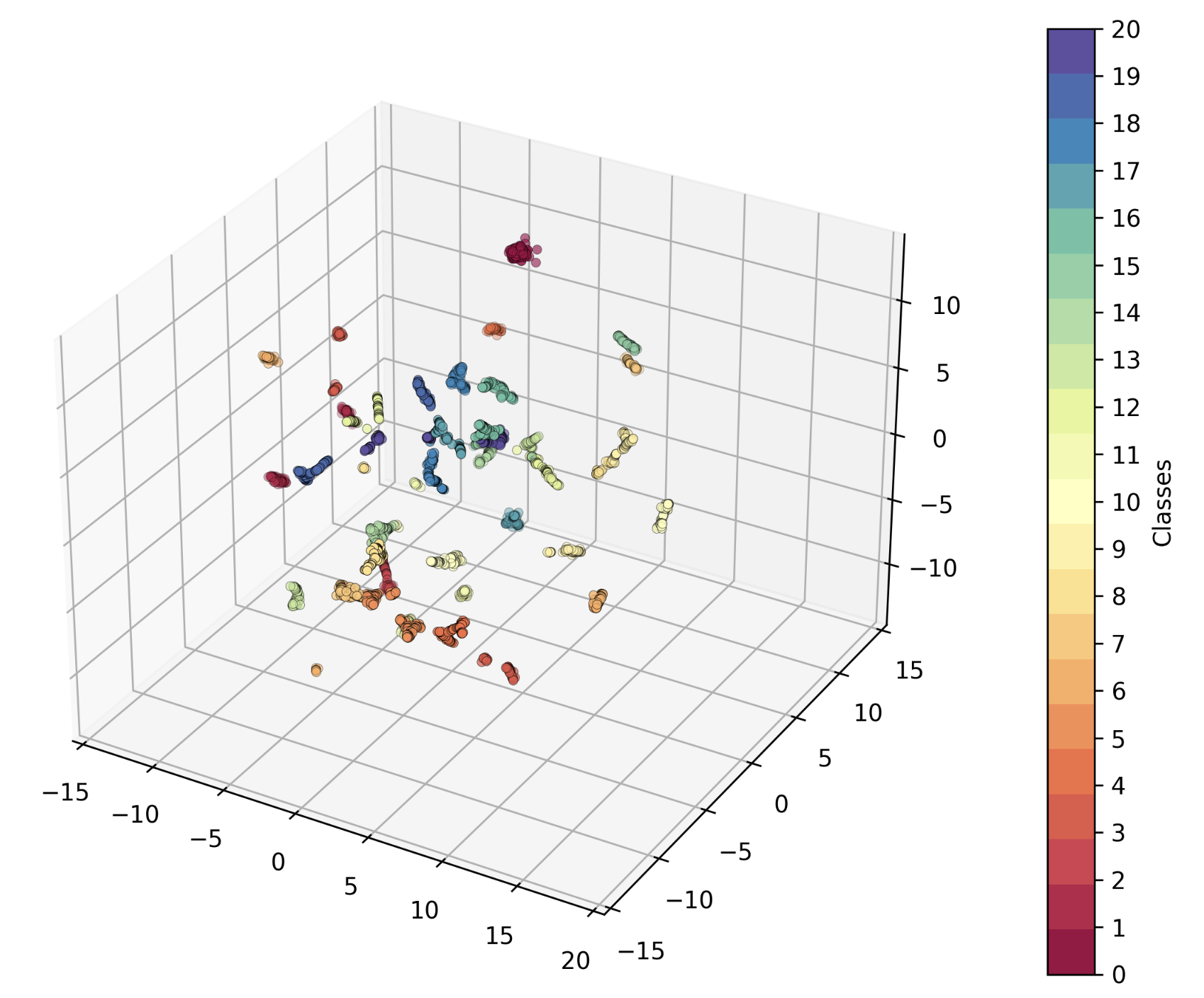}
        \caption{Original concatenated CSI features.}
        \label{fig:img6}
    \end{subfigure}
    \caption{The UMAP visualizations of both original and RSSI-weighted concatenated CSI features.}
    \label{fig:pic10}
\end{figure}
Then, the fused CSI features $\mathbf{F}_{p} \in \mathbb{R}^{C \times \frac{T_w}{2} \times \frac{S}{4}}$ can be expressed as
\begin{align}
    \mathbf{F}_{p} = (\mathbf{Q}^I_{1, p} \odot \mathbf{F}^D_{1, p}) \ || \ (\mathbf{Q}^I_{2, p} \odot  \mathbf{F}^D_{2, p}), 
\end{align}
where $\odot$ represents the element-wise multiplication operation, and $||$ represents the concatenation operation along time dimension. Such a RSSI-assisted fusion mechanism allows the fused CSI features to adaptively adjust the contribution of sensing features extracted from multi-view CSI data, which helps to improve the discrimination of fused features. To demonstrate the effectiveness of our RSSI-assisted fusion mechanism in improving the discrimination of fused CSI features, we examine the inter-class and intra-class variances for both the original concatenated CSI features $\mathbf{F}^{o}_{p}$ and the RSSI-weighted concatenated CSI features $\mathbf{F}_{p}$ using the testing dataset. Given the high dimensionality of these hidden features, we utilize UMAP \cite{mcinnes2018umap-software} to reduce the feature dimensions, enabling efficient computation of inter-class and intra-class variances. Fig.~\ref{fig:pic10} shows the UMAP visualizations of both original and RSSI-weighted concatenated CSI features. We can see that the clusters in Fig.~\ref{fig:img5} are more compact, with clearer separations between different classes. Fig.~\ref{fig:img5} also has denser clusters (e.g., see class 10), suggesting that features within the same class are more similar to each other.

Denote $\mathbf{U}^{o}_{p}$ and $\mathbf{U}_{p}$ as the features generated after UMAP dimemsionality reduction, the inter-class variance for class $l$ can be computed as
\begin{align}
    V_{w, l} = \sum \frac{1}{(|\mathcal{D}^t_l| - 1)} \sum_{p=1}^{|\mathcal{D}^t_l|} (\mathbf{U}_{p, l} - \boldsymbol{\mu}_l)^2, \\
    V^o_{w, l} = \sum \frac{1}{(|\mathcal{D}^t_l| - 1)} \sum_{p=1}^{|\mathcal{D}^t_l|} (\mathbf{U}^{o}_{p, l} - \boldsymbol{\mu}^{o}_l)^2,    
\end{align}
where $|\mathcal{D}^t_l|$ is the total testing samples of class $l$, $\boldsymbol{\mu}_l \triangleq \frac{1}{|\mathcal{D}^t_l|} \sum_{p=1}^{|\mathcal{D}^t_l|} \mathbf{U}_{p, l}$ and $\boldsymbol{\mu}^{o}_l \triangleq \frac{1}{|\mathcal{D}^t_l|} \sum_{p=1}^{|\mathcal{D}^t_l|} \mathbf{U}_{p, l}$ are the class mean (class centroid) of the $l$-th class for $\mathbf{U}^{o}_{p}$ and $\mathbf{U}_{p}$ respectively. And the intra-class variance can be computed as \begin{figure}[htbp]
    \centering
    \includegraphics[width=1.0\linewidth]{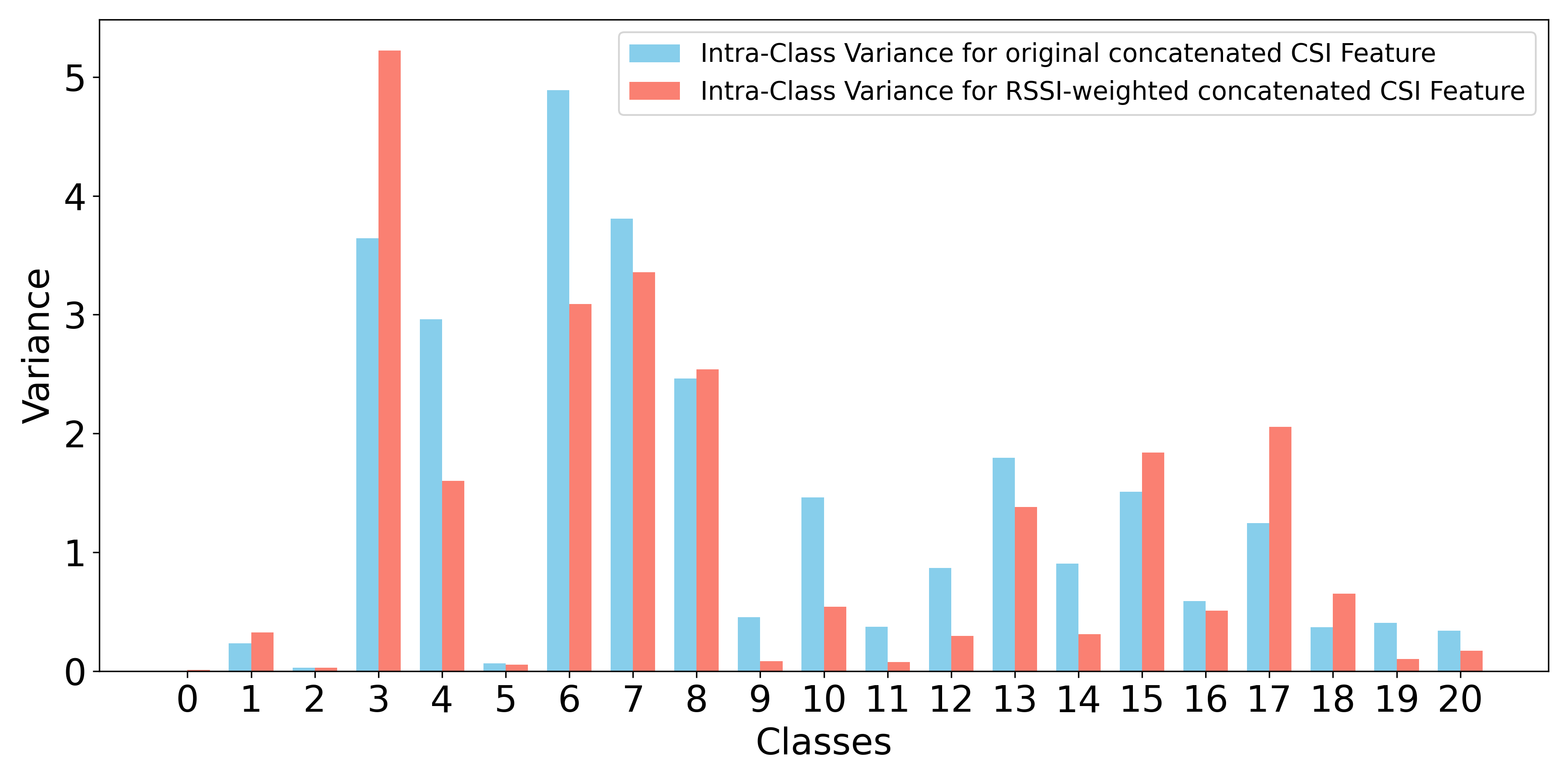}
    \caption{The intra-class variance for both the original and RSSI-weighted concatenated CSI features after dimensionality reduction using UMAP.}
    \label{fig:pic11}
\end{figure}\begin{table}[htbp]
\centering
\caption{The inter-class variance and average intra-class variance for both the original and RSSI-weighted concatenated CSI features after dimensionality reduction using UMAP.}
\label{tab:tab4}
\resizebox{1.0\linewidth}{!}{
\begin{tabular}{ccc}
\toprule 
\textbf{Feature Type} & \textbf{Average Intra-class Variance} & \textbf{Inter-class Variance}\\\midrule
Original concatenated CSI Feature & 1.35 & 1.74\\ 
RSSI-weighted concatenated CSI Feature & \textbf{1.16} & \textbf{1.94}\\
\bottomrule
\end{tabular}
}
\end{table}
\begin{align}
    V_b = \sum \frac{1}{L - 1} \sum_{l=1}^L (\boldsymbol{\mu}_l - \boldsymbol{\mu})^2, \\
    V^{o}_{b} = \sum \frac{1}{L - 1} \sum_{l=1}^L (\boldsymbol{\mu}^{o}_l - \boldsymbol{\mu}^o)^2,
\end{align}
where $L$ represents the total number of passenger number categories, $\boldsymbol{\mu} \triangleq \frac{1}{|\mathcal{D}^t|} \sum_{p=1}^{|\mathcal{D}^t|} \mathbf{U}_{p}$ and $\boldsymbol{\mu}^o \triangleq \frac{1}{|\mathcal{D}^t|} \sum_{p=1}^{|\mathcal{D}^t|} \mathbf{U}^{o}_{p}$ are the global mean (global centroid). Fig.~\ref{fig:pic11} and Table~\ref{tab:tab4} present the intra-class variance, inter-class variance, and average intra-class variance for both the original and RSSI-weighted concatenated CSI features, following dimensionality reduction with UMAP. Overall, the combination of lower intra-class variance and higher inter-class variance in RSSI-weighted concatenated CSI features suggests an enhancement in the feature space structure. Internally, the classes are more compact, while externally, they are more distinguishable, which is advantageous for classification tasks. Notably, Fig.~\ref{fig:pic11} indicates that RSSI reweighting can cause increased intra-class variance in some classes, such as class 3. Nevertheless, this effect can be alleviated by further backbone training.

The fused CSI features $\mathbf{F}_{p}$ are then fed into the CNN backbone $f^s_{\boldsymbol{\alpha}}(\cdot)$ with trainable parameters $\boldsymbol{\alpha}$ for generating the predicted probability distribution of the number of passenger. Finally, we adopt the class with the highest probability as the prediction result, i.e., $\hat{y}_p \triangleq f_{\text{RSSI-CSI}}\left(\mathbf{D}_{1, p}, \mathbf{I}_{1, p}, \mathbf{D}_{2, p}, \mathbf{I}_{2, p} ; \boldsymbol{\Sigma} \right)$ with $\boldsymbol{\Sigma} \triangleq \left\{ \boldsymbol{\theta},\boldsymbol{\beta},\boldsymbol{\gamma}, \boldsymbol{\alpha}\right\}$ encapsulating all the trainable parameters in the proposed framework.

\subsection{Training Procedure}
The training procedure involves optimizing the feature extractors in both receivers, the feature fusion module, and the CNN backbone to minimize the cross-entropy loss for accurate passenger counting, which can be expressed as:
\begin{equation}
\label{eqn_8}
\mathcal{L} \triangleq - \frac{1}{|\mathcal{D}|}\sum_{p=1}^{|\mathcal{D}|} y_p \log \left(\hat{y}_p\right),
\end{equation}
where $|\mathcal{D}|$ denotes the total amount of training samples collected from single Wi-Fi Tx-Rx pair and  $y_p$ is the ground truth label of $p$-th CSI training segment.

\section{Experimental Evaluation}
In this section, we evaluate the proposed Wi-Fi-based passenger counting system using a real-world Wi-Fi CSI dataset collected on a double-decker bus in Hong Kong.

\subsection{Experimental Setup}
The Wi-Fi CSI dataset for passenger counting is collected on a double-decker bus with up to 20 passengers engaged in static activities (e.g., sitting, talking, sleeping). 
For each of the 21 counting scenarios, the collection of CSI data entailed a duration of $2$ minutes, during which, $K=120,000$ packets were generated for each Wi-Fi Tx-Rx pair. The number of subcarriers was configured to $S=256$. 

The Wi-Fi CSI data were captured via the "tcpdump" interface and stored in packet capture (PCAP) files. Nexcsi\cite{nexcsi} was adopted to interpret the PCAP files, which produced a $12,000\times 256$ raw Wi-Fi CSI matrix for each PCAP file. 
Following the data pre-processing steps in Sec. \ref{sec:datapreprocessing}, the $14$ null subcarriers in raw Wi-Fi CSI matrix are removed, yielding a $12,000 \times 242$ Wi-Fi CSI matrix. By setting $T_w = 300$ and $T_s=150$ for data segmentation, we obtain $790$ CSI data segments with dimemsion $300 \times 242$ for each Wi-Fi Tx-Rx pair per passenger number category. The number of CSI data segments used for training and testing are configured to $13,272$ and $3,318$, respectively. 

For the CNN backbone at the edge server, the MobileOne-S4 model~\cite{vasu2023mobileone} is selected due to its high efficiency and high accuracy. The classification models are trained for 100 epochs with the AdamW optimizer. We configure the batch size, initial learning rate, and weight decay to $64$, $10^{-3}$, and $10^{-4}$, respectively. In addition, CosineAnnealingLR scheduler is applied to adjust the learning rate, and the cross-entropy loss function with label smoothing of $0.1$ is used as the learning objective function.

\subsection{Performance Analysis}
In this section, we provide a comprehensive performance evaluation of our system under a mimic real-world passenger counting scenario.

To highlight the advantages of the proposed method, we also compare our approach against several baselines for both single-receiver sensing and multi-receiver cooperative sensing: 
\begin{enumerate}
    \item \textbf{Receiver 1}: This baseline uses the data collected from Receiver 1 to train the model. 
    \item \textbf{Receiver 2}: In this baseline, the model is trained using data gathered from Receiver 2.   
    \item \textbf{Direct Probability Averaging (Direct Prob Avg)}: In this baseline approach, the final decision is determined by directly averaging the probability outputs from two models 
    \item \textbf{Re-weighted CSI Probability Averaging (Re-weighted CSI Prob Avg)}: This approach first calculates the RSSI weights, $\hat{\mathbf{I}}_{n, p} := \frac{\mathbf{I}_{n, p}}{\sum_{n = 1}^N \mathbf{I}_{n, p}}$, to generate re-weighted CSI data, $\hat{\mathbf{D}}_{n, p} \triangleq \hat{\mathbf{I}}_{n, p} \odot \mathbf{D}_{n, p}$, before feeding it into the models. It then averages the probability outputs from the two models to make the final decision. 
    \item \textbf{CSI Feature Concatenation Training}: Here, the CSI features extracted from the feature extractors of all Wi-Fi Tx-Rx pairs are directly concatenated and then fed into the CNN backbone for final decision-making. 
\end{enumerate}
Our approach is similar to the fourth baseline, which also introduces additional RSSI sensing information, but enhances it by adaptively adjusting the RSSI weights through a learning-based method, without employing a probability ensemble. Through latent feature fusion, CSI and RSSI data collected from multiple receivers are encouraged to align in a common latent space, allowing an adaptive merging of informative features. We concatenate RSSI features along the time dimension, instead of the channel dimension, and input them into a convolution layer with kernel size and stride matching the time length for each RSSI feature, without padding. This design enables our fusion module to extend to scenarios with more receivers ($N > 2$) without retraining. Sensing performance is evaluated using accuracy and F1-score. The experiments are conducted using a fixed seed to ensure reproducibility of the results. And we report the final results by averaging the results over the last ten training epochs to avoid getting a occasional result. The methods with the best performance are highlighted in \textbf{bold} and the second-best results are \uline{underlined}.

\begin{figure}[htbp]
    \centering
    \begin{subfigure}[htbp]{0.49\linewidth}
        \includegraphics[width=\linewidth]{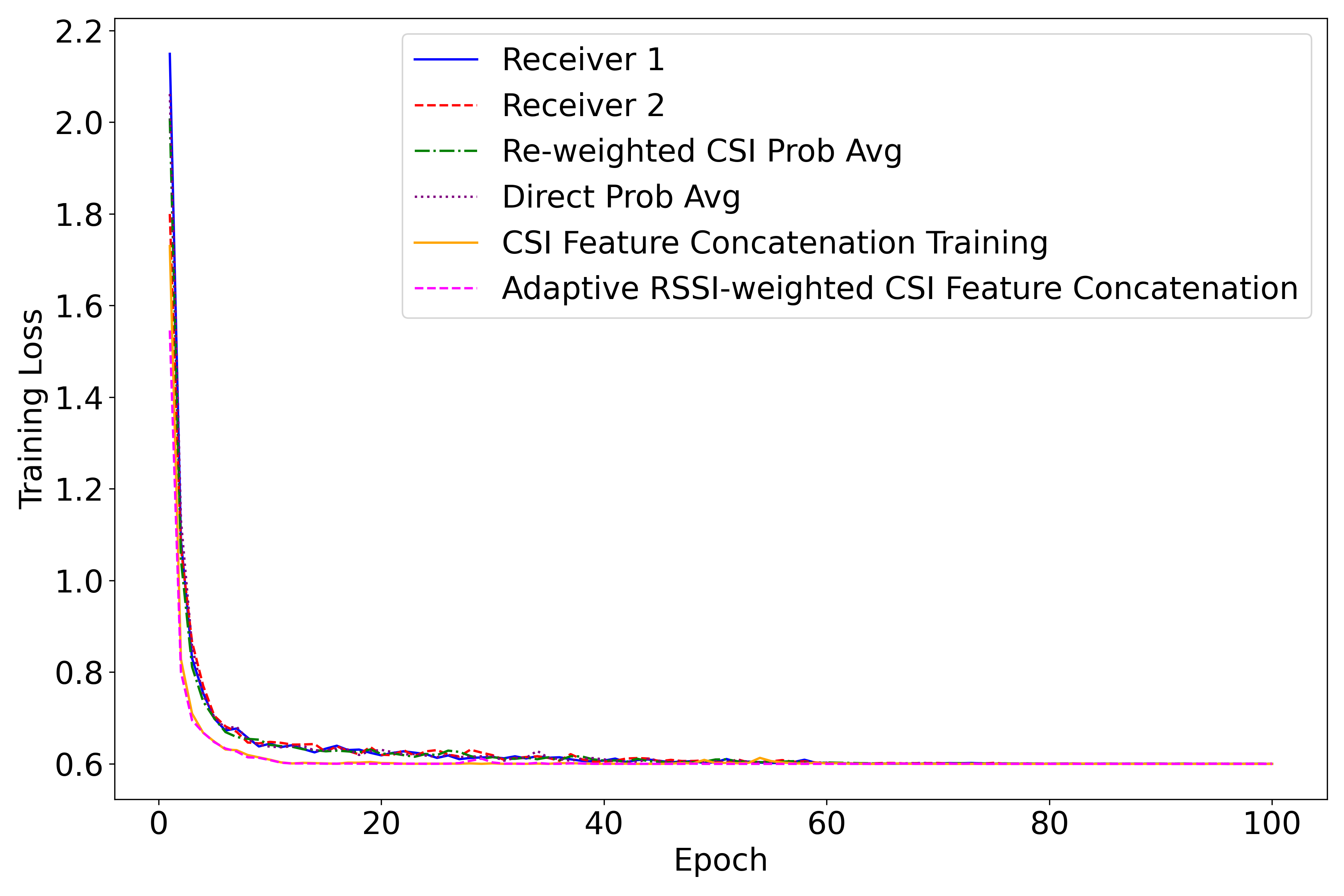}
        \caption{Training loss}
        \label{fig:img1}
    \end{subfigure}
    \begin{subfigure}[htbp]{0.49\linewidth}
        \includegraphics[width=\linewidth]{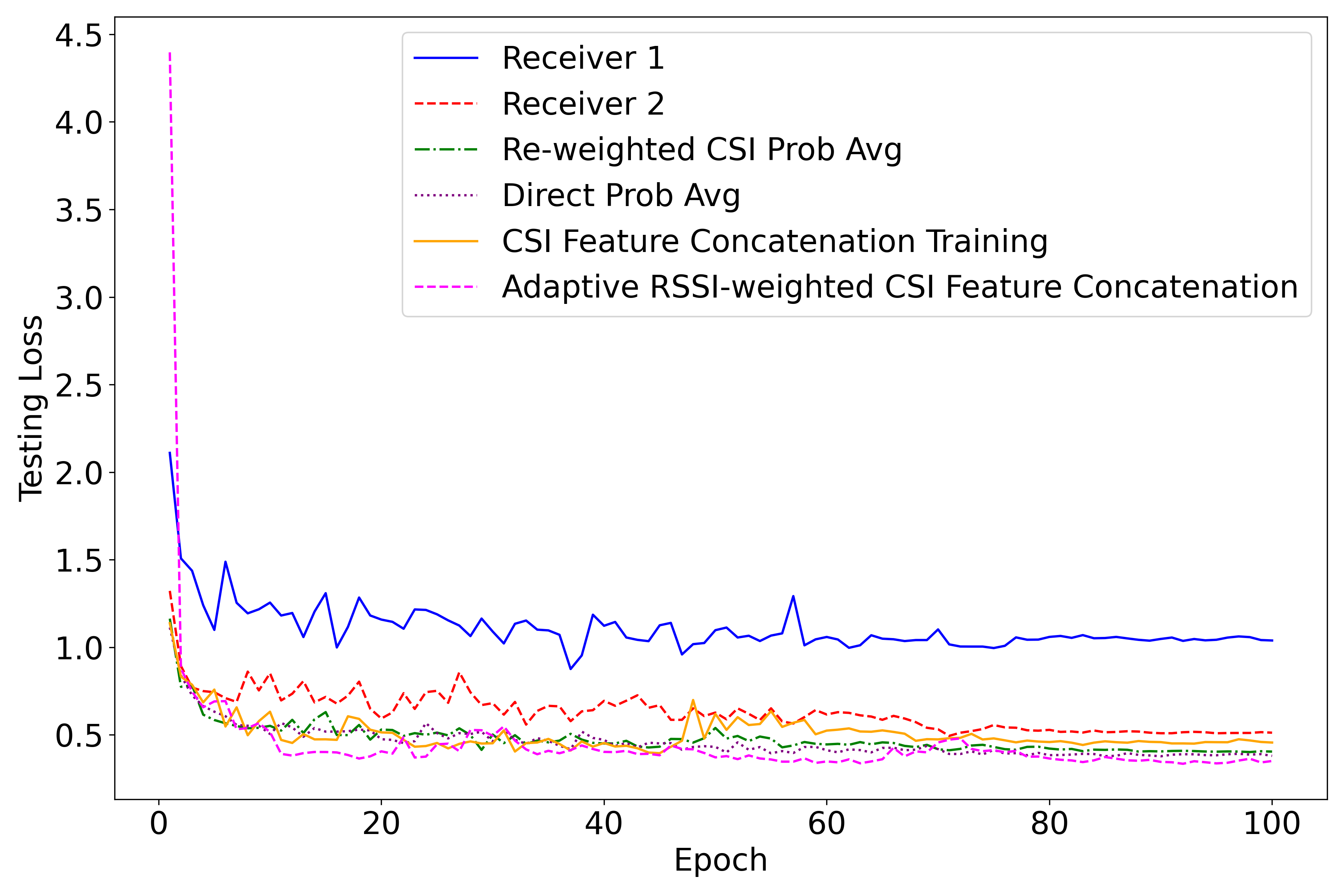}
        \caption{Testing loss}
        \label{fig:img2}
    \end{subfigure}
  \begin{subfigure}[htbp]{0.49\linewidth}
        \includegraphics[width=\linewidth]{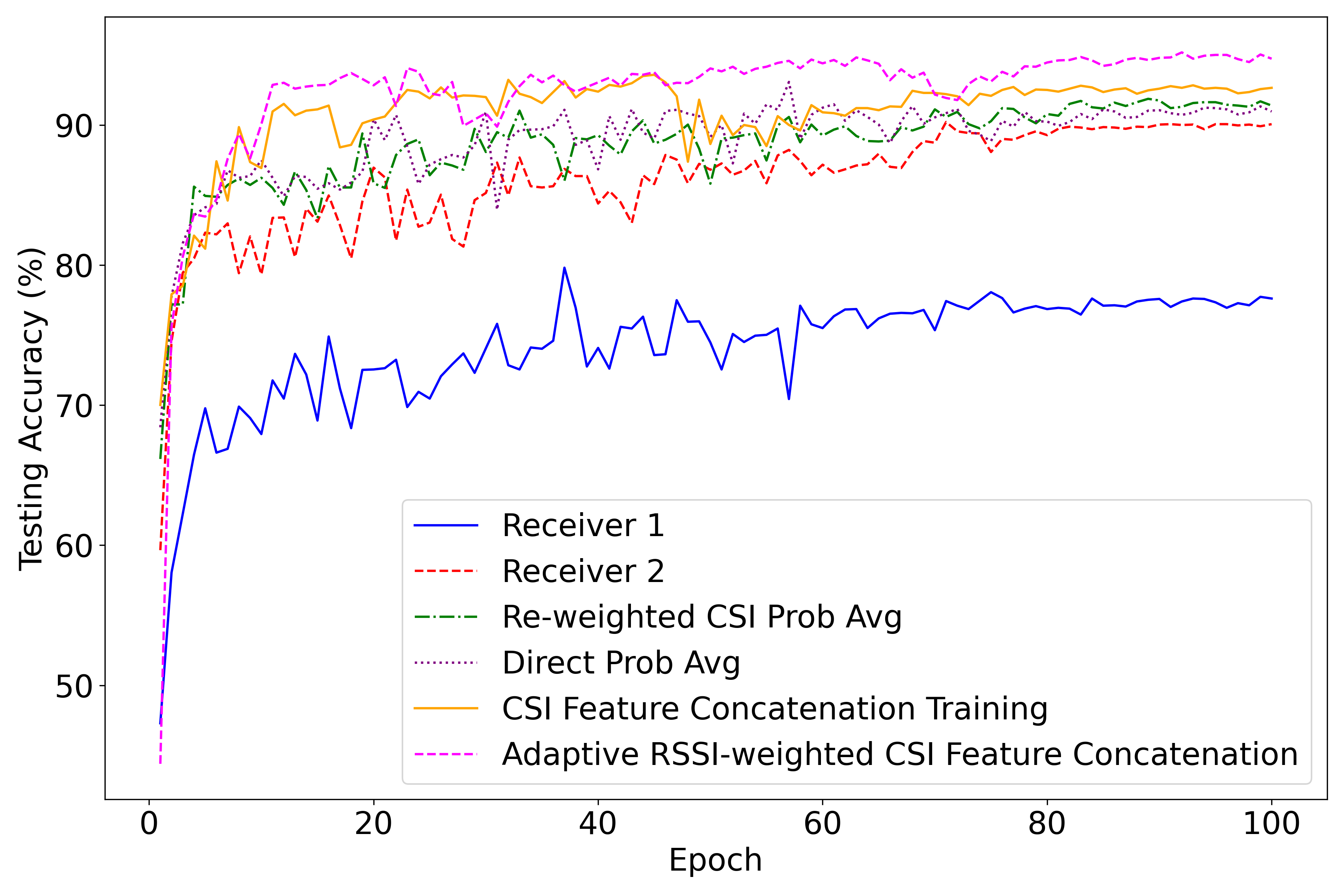}
        \caption{Testing accuracy}
        \label{fig:img3}
    \end{subfigure}
    \begin{subfigure}[htbp]{0.49\linewidth}
        \includegraphics[width=\linewidth]{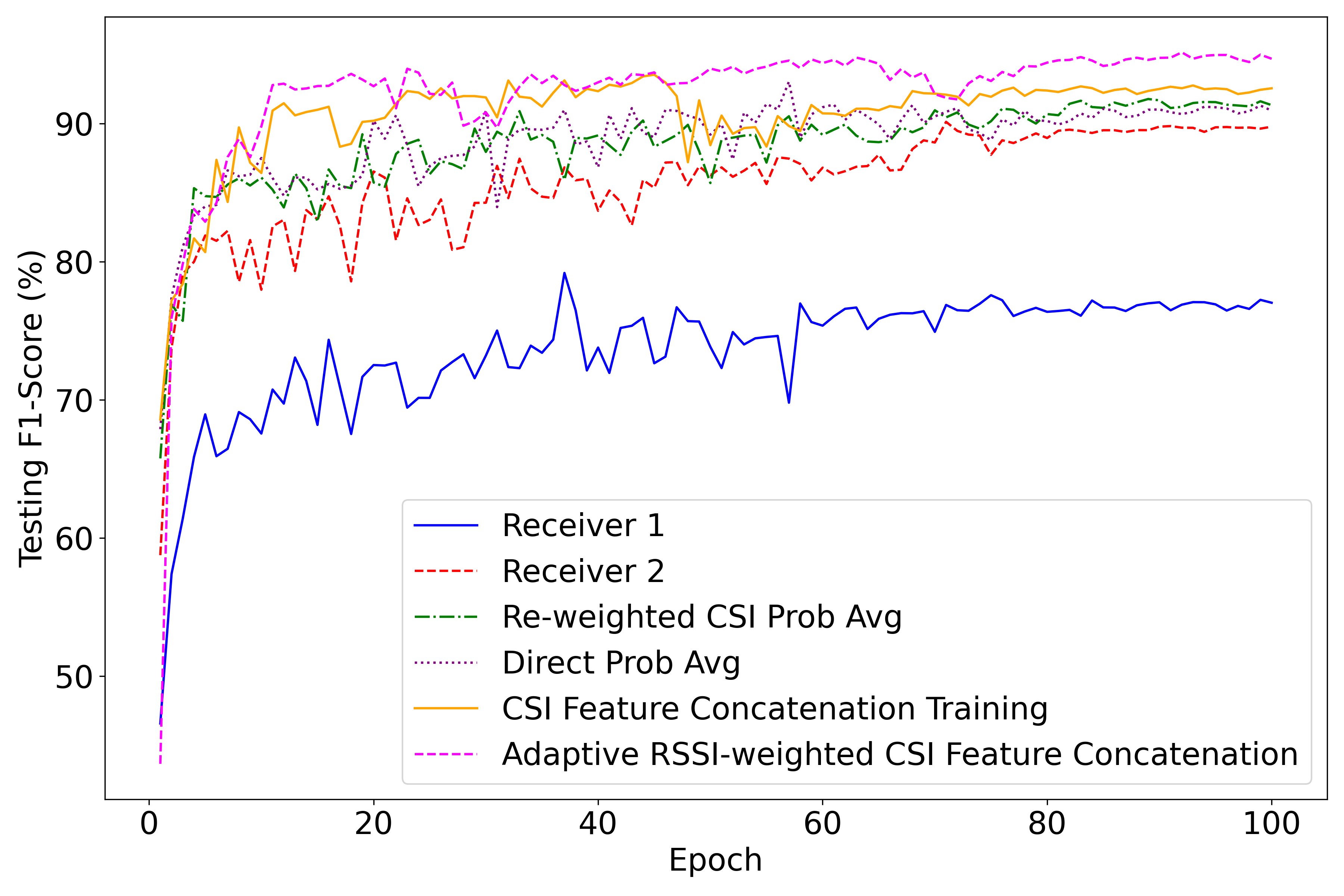}
        \caption{Testing F1-score}
        \label{fig:img4}
    \end{subfigure}
    \caption{The training and testing metric curves for the baselines and our proposed method under a specific training seed.}
    \label{fig:pic7}
\end{figure}

Fig.~\ref{fig:pic7} presents the convergence curves of training and testing metrics for both the baseline models and our proposed system, indicating that all methods have been properly trained after 100 epochs. Table~\ref{tab:tab3} shows the overall performance of using Wi-Fi CSI data with either single Wi-Fi Tx-Rx pair or two Wi-Fi Tx-Rx pairs for passenger counting, where the counting performance of receiver installed in the front of the upper decker (Receiver 2) is better than that installed at the rear of the upper decker (Receiver 1). The underlying reason behind it may be comprised of two parts according to the overall layout shown in Fig.~\ref{fig:pic1}: (1) two pingers are installed in the front of the upper decker, which allow more updated CSI data being collected via Receiver 2; (2) the layout for the front of the upper decker is asymmetries and therefore, result in more diverse CSI data being generated and collected by Receiver 2. Our proposed method surpasses the performance with single Wi-Fi Tx-Rx pair by up to 17.5\% in accuracy and 17.97\% in F1-score, which demonstrates the effectiveness of CSI data cooperative approaches introduced in our proposed system. Since our method integrates extra RSSI sensing information and adaptively adjust the RSSI weight for CSI feature fusion, it achieves performance 
improvement with up to 3.87\% in accuracy and 3.85\% in F1-score when compared with those multi-receiver cooperative sensing baselines. Meanwhile, the overall counting performance of our proposed method is boosted to be above 94\%. Furthermore, by integrating RSSI information, the Re-weighted CSI Prob Avg baseline sees a marginal gain compared to the Direct Prob Avg baseline, emphasizing the advantage of enriching CSI data with RSSI information  

The computational overhead of the proposed method and baselines are also summarized in Table~\ref{tab:tab3}, where GFLOPs denoted as giga floating point of operations. It can be observed that the RSSI weight calculation of the proposed approach is lightweight and only increase about 0.01 GFLOPs when comparing with CSI Feature Concatenation Training method. 

\section{Conclusion}
In this paper, we developed a passenger counting system for double-decker buses that leverages multiple Wi-Fi receivers to enhance the counting performance. Specifically, a novel RSSI-assisted cooperative couting method is developed, which exploits RSSI to provide extra sensing information to the CSI data. First, RSSI and CSI data are extracted as informative features for transmitting to the edge server. Then a RSSI-assisted feature fusion module is proposed to adaptively integrated RSSI information into CSI features for augmenting the fused CSI feature. Experimental results show that our proposed system outperforms other multi-receiver cooperative sensing baselines in counting performance, which achieves at least 2.27\% accuracy and 2.34\% F1-score performance boost in a two-receiver setup. \begin{table}[htbp]
\centering
\caption{Overall performance and computational overhead of both single-receiver sensing and two-receiver cooperative sensing.}
\label{tab:tab3}
\resizebox{1.0\linewidth}{!}{
\begin{tabular}{cccc}
\toprule
\textbf{Method} & \textbf{Accuracy} & \textbf{F1-score} & \textbf{GFLOPs}
\\\midrule
Receiver 1 & 77.36 & 76.86 & \multirow{2}{*}{\textbf{1.72}} \\
Receiver 2 & 89.98 & 89.69 & \\\midrule
Direct Prob Avg & 90.99 & 90.98 & \multirow{2}{*}{3.44}
 \\
Re-weighted CSI Prob Avg & 91.45 & 91.39 & 
\\\midrule
CSI Feature Concatenation Training & \uline{92.59} & \uline{92.49} & \uline{3.40}
\\\midrule
Adaptive RSSI-weighted CSI Feature Concatenation & \textbf{94.86} & \textbf{94.83} & 3.41 \\\bottomrule
\end{tabular}
}
\end{table}The proposed system can incorporate few-shot learning for environment adaptation, and integrate with federated split learning for cross domain Wi-Fi sensing. Future work will explore semi-supervised learning to reduce extensive labeling needs and analyze CSI data collected from different positions to enhance sensing performance.

\normalem
\bibliographystyle{IEEEtran}

\end{document}